\documentclass[amsmath,amssymb,11pt,superscriptaddress,reprint, preprintnumbers, notitlepage,aps,prl,twocolumn]{revtex4-1}

\pdfoutput=1 
\usepackage[utf8]{inputenc}
\usepackage[english]{babel}
\usepackage{amsmath}
\usepackage{graphicx}
\usepackage{dcolumn}
\usepackage{pbox}
\usepackage{amssymb}
\usepackage{epsfig}
\usepackage[dvipsnames]{xcolor}
\usepackage{slashed}
\usepackage{amssymb}
\usepackage{ mathrsfs }
\usepackage{color}
\usepackage[font=small]{caption}
\usepackage[font=small]{subcaption}
\usepackage{url}
\definecolor{MyDarkBlue}{rgb}{0.1, 0.1, 0.8} 
\definecolor{SBlue}{rgb}{0.2, 0.4, 0.7} 
\definecolor{MyLightBlue}{rgb}{0.22,0.51,0.9}
\definecolor{MyGreen}{rgb}{0.0, 0.5, 0.0}
\definecolor{BrickRed}{rgb}{0.8, 0.25, 0.33}
\RequirePackage{hyperref}
\hypersetup{colorlinks, citecolor=SBlue,linkcolor=MyDarkBlue, urlcolor=PineGreen}

\makeatletter

\makeatletter
\renewcommand\@makecaption[2]{%
  \par
  \vskip\abovecaptionskip
  \begingroup
  
   \small\rmfamily
    \begingroup
     \samepage
     \flushing
     \let\footnote\@footnotemark@gobble
     \@make@capt@title{#1}{#2}\par
    \endgroup
  \endgroup
  \vskip\belowcaptionskip
}
\makeatother

\DeclareUnicodeCharacter{2212}{-}
\begin{document}

\title{\vspace{1cm}\Large 
Muon \boldmath${g-2}$ Anomaly and Neutrino Magnetic Moments
}

\author{\bf K.S. Babu}
\email[E-mail: ]{babu@okstate.edu}
\affiliation{Department of Physics, Oklahoma State University, Stillwater, OK, 74078, USA}    

\author{\bf Sudip Jana}
\email[E-mail:]{sudip.jana@mpi-hd.mpg.de}
\affiliation{Max-Planck-Institut f{\"u}r Kernphysik, Saupfercheckweg 1, 69117 Heidelberg, Germany}

\author{\bf Manfred Lindner}
\email[E-mail:]{lindner@mpi-hd.mpg.de}
\affiliation{Max-Planck-Institut f{\"u}r Kernphysik, Saupfercheckweg 1, 69117 Heidelberg, Germany}

\author{\bf Vishnu P.K.}
\email[E-mail:]{ vipadma@okstate.edu}
\affiliation{Department of Physics, Oklahoma State University, Stillwater, OK, 74078, USA}

\begin{abstract}
We show that a unified framework based on an $SU(2)_H$ horizontal symmetry which generates a naturally large neutrino transition magnetic moment and explains the XENON1T electron recoil excess also predicts a positive shift in the muon anomalous magnetic moment. This shift is of the right magnitude to be consistent with the Brookhaven measurement as well as the recent Fermilab measurement of the muon $g-2$.  A relatively light neutral scalar from a Higgs doublet with mass near 100 GeV contributes to muon $g-2$, while its charged partner induces the neutrino magnetic moment. In contrast to other multi-scalar theories, in the model presented here there is no freedom to control the sign and strength of the muon $g-2$ contribution.  We analyze the collider tests of this framework and find that the HL-LHC can probe the entire parameter space of these models.

\noindent 
\end{abstract}

\maketitle

\textbf{\emph{Introduction}.--} 
There has been considerable interest in understanding the long-standing discrepancy between the measured and predicted values of the  anomalous magnetic moment of the muon, $a_\mu$. The Brookhaven Muon g-2 collaboration has measured it to be $a_\mu({\rm BNL})=116 592 089(63)\times 10^{-11}$ two decades ago \cite{Bennett:2006fi}, while theoretical predictions find it to be $a_\mu ({\rm theory})=116 591 810(43)\times 10^{-11}$ \cite{Aoyama:2020ynm}.  Taken at face value the difference, $\Delta a_\mu = a_\mu({\rm experiment})- a_\mu({\rm theory}) \simeq  279 \times 10^{-11}$, is a 3.7 sigma discrepancy, which may indicate new physics lurking around or below the TeV scale.  Very recently the Fermilab Muon g-2 collaboration \cite{Abi:2021gix} has announced their findings, which measures it to be $a_\mu({\rm FNAL})=116 592 040 (54)\times 10^{-11}$  which confirms the Brookhaven measurement and increases the significance of the discrepancy to the level of 4.2 sigma. These results make the motivations for new physics explanation more compelling.  

\color{black} The purpose of this paper is draw a connection between new physics contributions to $a_\mu$ and a possible neutrino transition magnetic moment $\mu_{\nu_\mu \nu_e}$. A sizable neutrino magnetic moment has been suggested as a possible explanation for the excess in electron recoil events observed in the $(1 - 7)$ keV recoil energy range by the XENON1T collaboration recently \cite{Aprile:2020tmw}.  The neutrino magnetic moment needed to explain this excess lies in the range of $(1.6 - 2.4) \times 10^{-11} \mu_B$, where $\mu_B$ stands for the electron Bohr magneton \cite{Babu:2020ivd}. Such a value would require new physics to exist around the TeV scale. As we shall show in this paper, models that induce neutrino magnetic moments, while maintaining their small masses naturally, also predict observable shifts in the muon anomalous magnetic moment.  We focus on a specific class of models based on an $SU(2)_H$ horizontal symmetry (or family symmetry) acting on the electron and muon families that naturally leads to a large neutrino magnetic moment \cite{Voloshin:1987qy,Babu:1989wn}.  We find that within this class of models, an explanation of the XENON1T excess will necessarily lead to {\it a positive contribution} to $\Delta a_\mu$, which lies neatly within the Brookhaven  and the recent Fermilab measurements of $a_\mu$ \cite{Abi:2021gix}.  \color{black} This class of models is thus in accordance with Occam's razor, explaining both anomalies in terms of the same new physics.

\textbf{\emph{Model}.--} 
We focus on models that can generate sizable transition magnetic moments for the neutrinos while keeping their masses naturally small.  Models which generate a neutrino transition magnetic moment of order $10^{-11}\mu_B$ would typically also induce an unacceptably large neutrino mass of order 0.1 MeV.  An $SU(2)_H$ family symmetry acting on the electron and muon families can resolve this conundrum as it decouples the neutrino magnetic moment from the neutrino mass \cite{Babu:1989wn}.  This follows from the different Lorentz structures of the transition magnetic moment and mass operators. The magnetic moment operator $(\nu_e^T C \sigma_{\mu\nu}\nu_\mu) F^{\mu\nu}$, being antisymmetric in flavor is a singlet of $SU(2)_H$, while the mass operator $(\nu_e^T C \nu_\mu)$, being symmetric in flavor, belongs to $SU(2)_H$ triplet and  vanishes in the $SU(2)_H$ symmetric limit \cite{Voloshin:1987qy}. The class of models we study makes use of this symmetry argument.
New physics at the TeV scale is required to generate the magnetic moment of the desired magnitude.  This should violate lepton number, since the transition magnetic moment operator is a $\Delta L = 2$ operator.  This requirement naturally leads to the 
model of Ref. \cite{Babu:2020ivd}.  It is based on an $SU(2)_H$ extension of the Zee model of neutrino mass \cite{Zee:1980ai} involving a second Higgs doublet and a charged singlet $\eta^\pm$.  The extension incorporates the $SU(2)_H$ symmetry.

The class of models under study is based on the SM gauge group and an $approximate$ $SU(2)_H$ horizontal symmetry acting on the electron and muon families \cite{Babu:2020ivd,Babu:1989wn,Babu:1990wv}. The lepton fields transform under $SU(2)_L\times U(1)_Y \times SU(2)_H $ as follows: 
\begin{widetext}
\begin{equation}
     \psi_{L} =\begin{pmatrix} \nu_e & \nu_{\mu} \\ e & \mu  \end{pmatrix}_L\sim (2,-\dfrac{1}{2}) (2), \quad \psi_{R} =\left(e\quad \mu\right)_R\sim (1,-1)(2), \quad \psi_{3L} =\begin{pmatrix} \nu_{\tau} \\ \tau  \end{pmatrix}_L\sim (2,-\dfrac{1}{2})(1), \quad  \tau_R\sim (1,-1)(1).
\end{equation}
\end{widetext}
All quark fields transform as singlets of $SU(2)_H$ symmetry and will not be discussed further here. The scalar sector of the model contains the following fields:
\begin{widetext}
\begin{equation}
    \phi_S = \begin{pmatrix} \phi_S^+  \\ \phi_S^0  \end{pmatrix}\sim (2,\dfrac{1}{2})(1), \quad  {\Phi = \begin{pmatrix} \phi_{1}^{+} & \phi_{2}^{+} \\ \phi_{1}^{0} & \phi_{2}^{0}\end{pmatrix}}\sim(2,\dfrac{1}{2})(2), \quad \eta=\left(\eta_1^+ \quad \eta_2^+\right)\sim(1,1)(2).
\end{equation}
\end{widetext}
Here the $\phi_S$ field is the SM Higgs doublet which acquires a vacuum expectation value (VEV) $\langle \phi_S^0\rangle = v/\sqrt{2}$, with $v =246$ GeV.  The $\Phi$ and $\eta$ fields, which are doublets of $SU(2)_H$, are needed to generate transition magnetic moments for the neutrino.  Both fields are necessary in order to break lepton number. The neutral components of the $\Phi$ field are assumed to not acquire any VEV.
With this particle content, in the $SU(2)_H$ symmetric limit, the leptonic Yukawa couplings of the model are given by the Lagrangian \cite{Babu:2020ivd} 
\begin{eqnarray}
\mathcal{L}_{\rm Yuk} &=& h_{1} \operatorname{Tr}\left(\bar{\psi}_{L} \phi_{S} \psi_{R}\right)+h_{2} \bar{\psi}_{3 L} \phi_{S} \tau_{R}+h_{3} \bar{\psi}_{3 L} \Phi i \tau_{2} \psi_{R}^{T} \nonumber \\
&~&+ f \eta \tau_{2} \psi_{L}^{T} \tau_{2} C \psi_{3 L}+f^{\prime} \operatorname{Tr}\left(\bar{\psi}_{L} \Phi\right) \tau_{R} + H.c.
\label{Yuk1}
\end{eqnarray}
Eq. (\ref{Yuk1}) would lead to the relation $m_e = m_\mu$, which of course is unacceptable.  This is evaded by allowing for explicit but small $SU(2)_H$ breaking terms in the Lagrangian.  In particular, the following Yukawa couplings are necessary:
\begin{eqnarray}
{\cal L}_{\rm Yuk}' = \delta h_1 [(\overline{L}_e \phi_S e_R) - (\overline{L}_\mu \phi_S \mu_R)] + H.c.
\end{eqnarray}
Here $L_e^T = (e~ \nu_e)_L$, etc.
With this term included, $m_e \neq m_\mu$ is realized, since now we have $m_e = (h_1+ \delta h_1)v/\sqrt{2}$ and $m_\mu = (h_1-\delta h_1)v/\sqrt{2}$.  While the violation of $SU(2)_H$ is small, $\delta h_1 = (m_e - m_\mu)/(\sqrt{2} v) \simeq 3 \times 10^{-4}$, it is nevertheless significant in constraining the parameter space of the model, as we shall see. There are also other $SU(2)_H$ breaking terms in the Lagrangian of similar order, but these terms will not play any significant role in our discussions.

The scalar potential of the model contains $SU(2)_H$ symmetric terms of the form \cite{Babu:2020ivd}:
\begin{eqnarray}
V &\supset & m_\eta^2(|\eta_1|^2+|\eta_2|^2) + m_{\phi^+}^2 ( |\phi_1^+|^2 + |\phi_2^+|^2) \nonumber \\
&+&  m_{\phi^0}^2 (|\phi_1^0|^2 + |\phi_2^0|^2) + \{\mu\, \eta \,\Phi^{\dagger} i \tau_{2} \phi_{S}^{*} + H.c.\}
\label{potential}
\end{eqnarray}
The cubic term in the Eq.~(\ref{potential}) would lead to mixing between the charged scalars $\eta_i^+$ and $\phi_i^+$ (with $i=1,2$). The corresponding mass eigenstates are denoted as $h_i^+$ (with mass $m_{h^+}^2$) and $H_i^+$ (with mass $m_{H^+}^2$) and are given by 
\begin{align}
    & h_{i}^+=\cos{\alpha}\; \eta_i^+ + \sin{\alpha}\; \phi_i^+,\nonumber \\
    &  H_{i}^+= -\sin{\alpha}\; \eta_i^+ + \cos{\alpha}\; \phi_i^+,
\label{mixing}    
\end{align}
where the mixing angle $\alpha$ is defined as 
\begin{align}
&\tan 2\alpha= \frac{\sqrt{2}\mu v}{ m_{\eta}^2-m_{\phi^+}^2}. 
\end{align}
This mixing is essential for lepton number violation and the generation of magnetic moment of the neutrino. Note that the two copies of $h_i^+$ are degenerate, as are the two $H_i^+$ owing to $SU(2)_H$ symmetry.  

\textbf{\emph{Muon anomalous magnetic moment and neutrino transition magnetic moment}.--}
 \begin{figure}[htb!]
\includegraphics[height=3cm, width=0.52\textwidth]{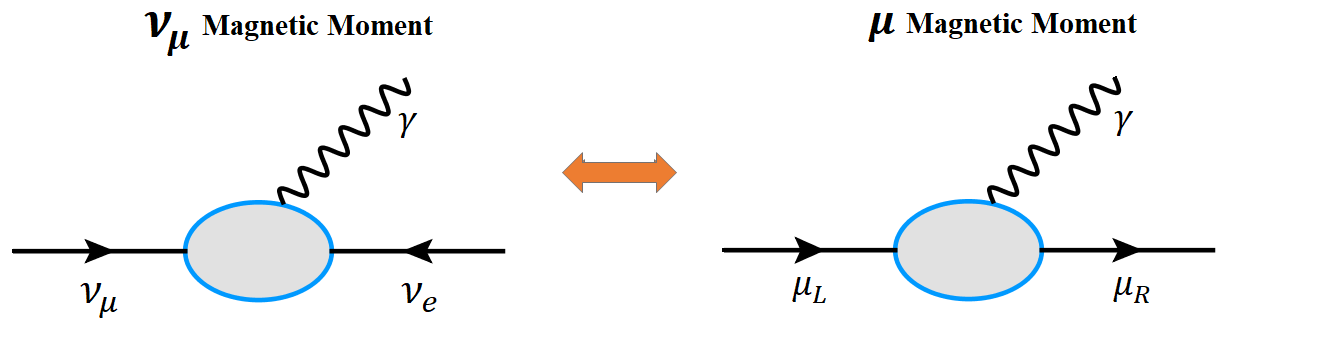}
\caption{New physics affecting muon anomalous magnetic moment as well as $\nu_{\mu}$-magnetic moment.  
} 
\label{feyn}
\end{figure}
 \begin{figure}[htb!]
\includegraphics[height=6.5cm, width=0.52\textwidth]{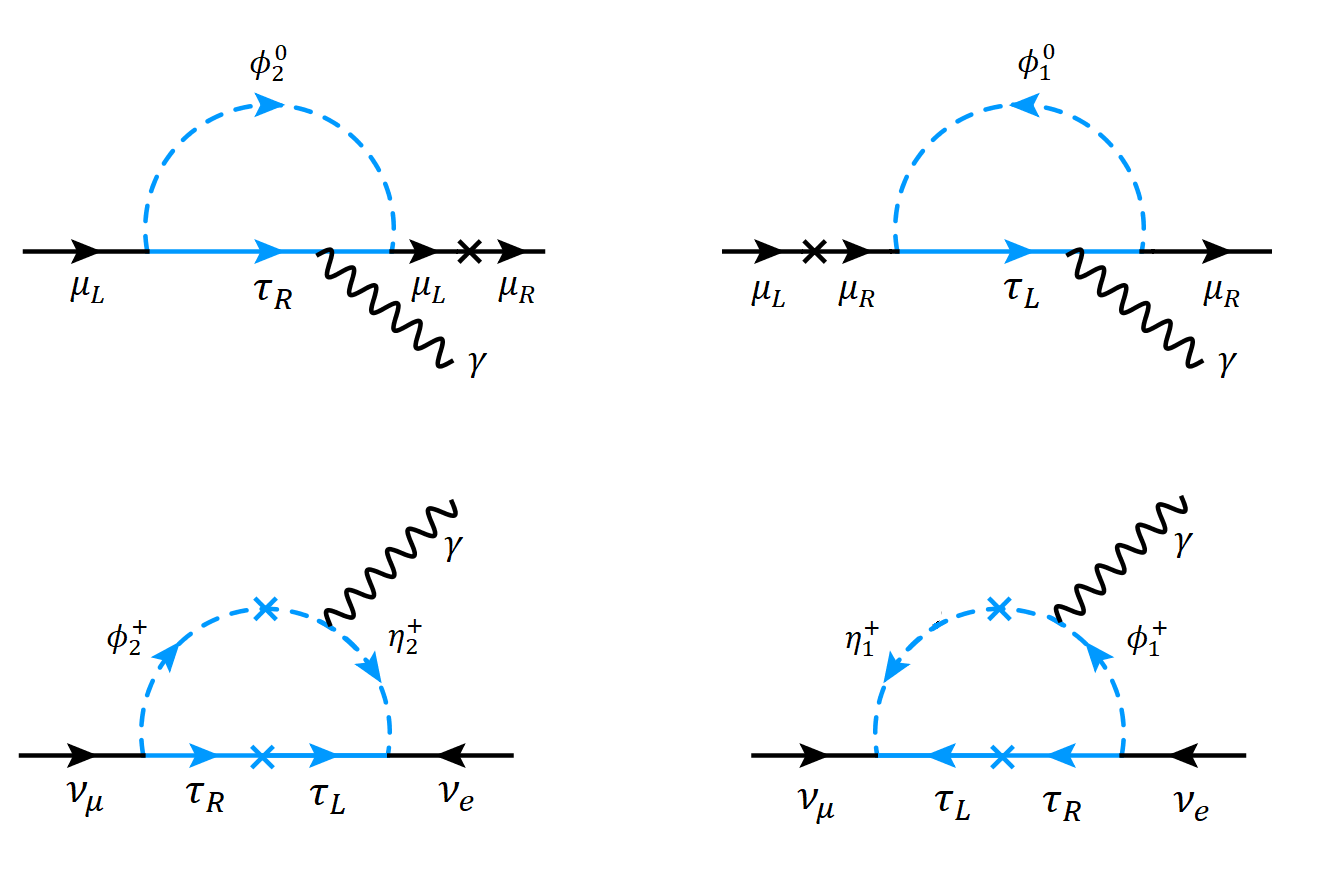}
\caption{The dominant contributions to $\Delta a_{\mu}$ (top) and neutrino magnetic moment (bottom) are shown.  For the neutrino magnetic moment, the outgoing photon can be emitted also from the $\tau$ lepton line. There are other diagrams, which are subleading.  
} \label{feynman}
\end{figure}
Fig. \ref{feyn} shows the connection between $\Delta a_\mu$ and $\mu_{\nu_\mu \nu_e}$ in a generic model.  Fig. \ref{feynman} has the explicit diagrams in the $SU(2)_H$ symmetric models.
We first focus on the new contributions to the muon anomalous magnetic moment within our framework. Both the neutral scalars and charged scalars present in the model contribute to $\Delta a_{\mu}$ via one-loop diagrams shown in Fig.~\ref{feynman}.
Since chirality flip occurs on the external legs in these diagrams, the loop corrections mediated by the neutral scalars $\phi_1^0$ and $\phi_2^0$ contribute positively to $a_\mu$, whereas the corrections from the charged scalars result in negative $a_\mu$. The full one-loop contributions to $\Delta a_\mu$ are given by \cite{Leveille:1977rc}
\begin{align}
\Delta a^{\varphi^0}_{\mu}&=
\frac{m_{\mu}^2}{16\pi^2}  
\left( |f'|^2+|h_3|^2 \right) F_{\varphi^0}[m_{\phi^0}], 
\label{amu0}
\\
\Delta a^{\varphi^+}_{\mu}&=
\frac{m_{\mu}^2}{16\pi^2} \left( |f\cos{\alpha} |^2+|h_3\sin{\alpha}|^2 \right)F_{\varphi^+}[m_{h^+}] 
\nonumber \\
& + \frac{m_{\mu}^2}{16\pi^2} \left(|\text{--}f\sin{\alpha} |^2+|h_3\cos{\alpha} |^2 \right)F_{\varphi^+}[m_{H^+}],
\label{amuplus}
\end{align}
where the loop functions are given by
\begin{align} 
&F_{\varphi^0}[m_{\varphi^0}]=
\int_0^1 dx \frac{x^2(1-x)}{m_{\mu}^2 x^2+m^2_{\varphi^0}(1-x)+x(m^2_{\tau}-m^2_{\mu})},
\\
&F_{\varphi^+}[m_{\varphi^+}]= \int_0^1 dx \frac{x^2(x-1)}{m_{\mu}^2x^2+x(m^2_{\varphi^+}-m_{\mu}^2)}.
\end{align}
 \begin{figure}[tb!]
\includegraphics[width=0.5\textwidth]{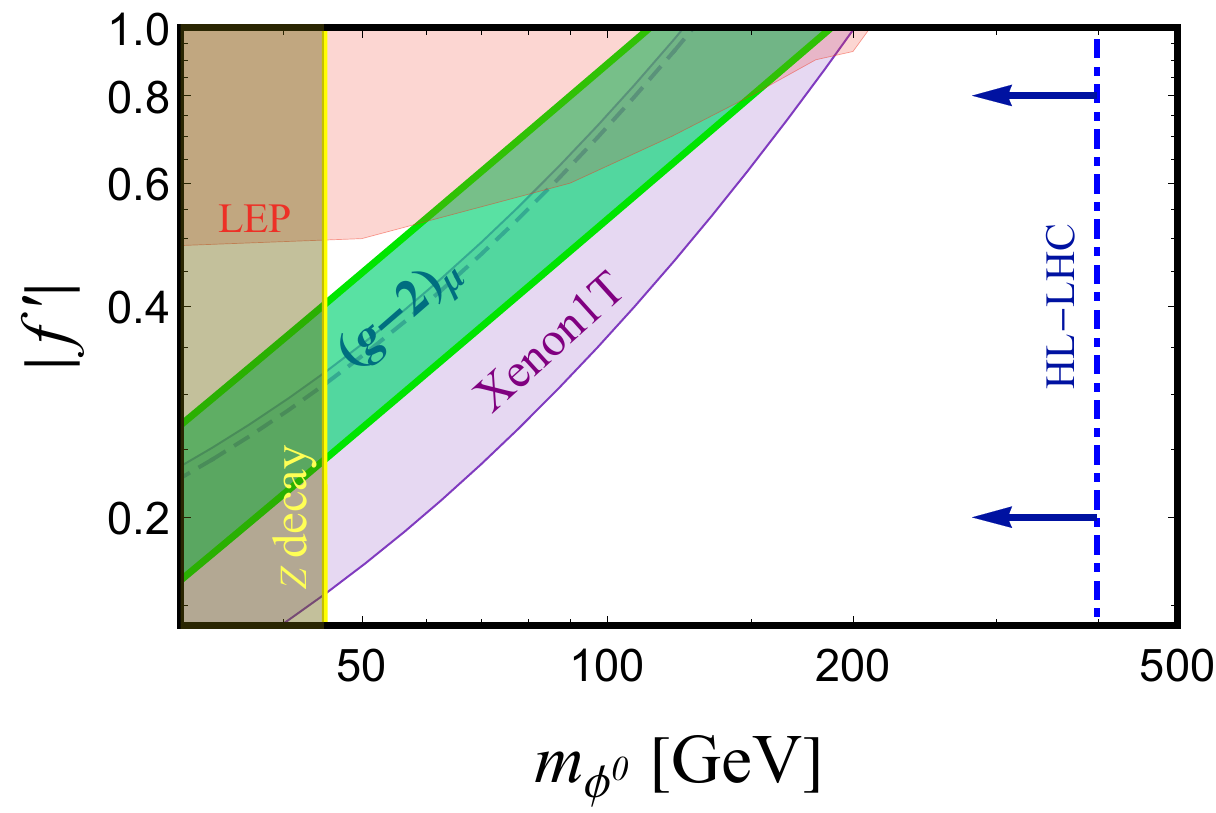}
\caption{{The $2\sigma$  allowed range (in green) for the muon anomalous magnetic moment measurement \cite{Abi:2021gix} at the Fermilab  in the $|f'|$-$m_{\phi^0}$ plane}. The region labelled XENON1T
explains the electron recoil excess at $90\%$ C.L \cite{Aprile:2020tmw}. The dashed line indicate the limit on neutrino magnetic
moment from the BOREXINO experiment \cite{Borexino:2017fbd}. The shaded regions denote the excluded parameter space from various experiments: the light pink shaded region from the LEP experiment \cite{LEP:2003aa} and the yellow shaded region from the $Z$-decay width measurements \cite{LEP:2003aa,Tanabashi:2018oca}. Here we have fixed $f$ = $10^{-2}$, $\sin \alpha$ = 0.35, $m_{h^+}=m_{\phi^0}+50$ GeV and $m_{H^+}=m_{h^+}+350$ GeV.
} \label{plota}
\end{figure}
\noindent Here $m_{\phi^0}$ is the common mass  $m_{\phi_1^0} = m_{\phi_2^0}$.  In Eq. (\ref{amu0}), the two terms proportional to $|f'|^2$ and $|h_3|^2$ arise from diagrams with $\phi_2^0$ and $\phi_1^0$ exchange respectively.  The two contributions in Eq. (\ref{amuplus}) arise from the exchange of $h^+$ and $H^+$.  

We now turn to the transition  magnetic moment of the neutrino in our framework and show its correlation with $\Delta a_\mu$. The diagrams generating sizeable $\mu_{\nu_\mu\nu_e}$ are shown in Fig. \ref{feynman}, bottom panel. Owing to the $SU(2)_H$ symmetry of the model, the two diagrams add in their contributions to the magnetic moment, while they subtract in their contributions to neutrino mass when the photon line is removed from these diagrams (for details, see Ref. \cite{Babu:2020ivd,Babu:1990wv}).  The resulting neutrino magnetic moment is given by \cite{Babu:2020ivd,Babu:1990wv}
 \begin{widetext}
\begin{equation}
\mu_{\nu_\mu\nu_e}=\frac{f f^{\prime}}{8 \pi^{2}} m_{\tau} \sin 2 \alpha\left[\frac{1}{m_{h^+}^{2}}\left\{\ln \frac{m_{h^+}^{2}}{m_{\tau}^{2}}-1\right\}-\frac{1}{m_{H^+}^{2}}\left\{\ln \frac{m_{H^+}^{2}}{m_{\tau}^{2}}-1\right\}\right]~.
\label{moment}
\end{equation}
 \end{widetext}
Comparing Eq. (\ref{moment}) with Eqs. (\ref{amu0}) and (\ref{amuplus}) we see that $\Delta a_\mu$ and $\mu_{\nu_\mu \nu_e}$ are dependent on the same set of parameters.  To see their correlation quantitatively, we have to take into account the various constraints that exist on these parameters, which we now address.

First of all, for $\mu_{\nu_\mu \nu_e}$ to be in the range of $(1.6-3.4) \times 10^{-11} \mu_B$ so that the XENON1T excess is explained, one needs $|f f' \sin2\alpha| \geq (1-8) \times 10^{-3}$, corresponding to $m_{h^+} = (100 - 300)$ GeV. The coupling $|f|$ can be constrained from the induced neutrino masses from the $SU(2)_H$ breaking effects.  The cancellation that occurs among the two diagrams in the bottom panel of Fig. \ref{feynman} in neutrino mass is valid only in the strict $SU(2)_H$ limit.  Since $m_e \neq m_\mu$ is necessary, one should examine neutrino masses including $SU(2)_H$ breaking effects. The resulting neutrino mass is given by \cite{Babu:2020ivd}
\begin{align}
m_\nu \simeq
\frac{ff'm_{\tau}\sin 2\alpha}{16\pi^2}
\left[ \frac{\delta m_{\eta}^2}{m_{\eta}^2}-\frac{\delta m_{\phi}^2}{m_{\phi}^2}+2(\delta \alpha) \cot{2\alpha} \text{ln}\frac{m_{\phi}^2}{m_{\eta}^2}  \right].
\label{nu}
\end{align}
Here $\delta m_{\eta}^2$ is the squared mass splitting between the two charged scalars $\eta_1^+$ and $\eta_2^+$ arising from $SU(2)_H$ breaking effects. Similarly, $\delta \alpha$ is the shift in the common mixing angle $\alpha$ that parametrizes $\eta_1^+-\phi_1^+$ mixing and $\eta_2^+-\phi_2^+$ mixing.  The mass splitting among charged scalars arising from $m_e \neq m_\mu$ is found to be 
\begin{align}
\delta m_{\eta}^2\simeq\frac{|f|^2(m_{\mu}^2-m_e^2)}{16\pi^2} 
\label{nu1}
\end{align}
which does not induce large neutrino mass, even for $|f| \sim 1$. (Eq. (\ref{nu1}) should replace Eq. (4.35) of Ref. \cite{Babu:2020ivd}, which is incorrect.) However, the coupling $|f|$ is constrained from the universality limits from $\tau$ decay.  In Ref. \cite{Babu:2020ivd} it was shown that the leptonic decay rate of the $\tau$ will receive additional contributions from the charged scalars leading to the modification $\Gamma(\tau \rightarrow e) = \Gamma_{\rm SM} \times (1+\epsilon_\tau)^2$, where
\begin{equation}
\epsilon_\tau = \frac{f^2}{g^2} m_W^2 \left(\frac{\cos^2\phi}{m_{h^+}^2} + \frac{\sin^2\phi}{m^2_{H^+}}  \right)~.
\end{equation}
A limit of $\epsilon_\tau \leq 0.004$ can be derived at 2 $\sigma$ from the constraint $g_\tau/g_\mu = 1.0011 \pm 0.0015$ \cite{Pich:2013lsa} obtained by comparing $\tau$ decay rate into muon with muon decay rate. This leads to the constraint $|f| \leq (0.05-0.25)$ for $h^+$ mass in the range of $(100-500)$ GeV.  
As a result, the charged scalar contributions to the muon anomalous magnetic moment turn out to be not significant.  

The coupling $h_3$ is constrained  from the induced tau neutrino mass within the model, which is given by
\begin{equation}
    m_{\nu_\tau} = \frac{h_3 f \sin 2\alpha}{32\pi^2} m_\mu {\rm ln}\left(\frac{m_{h^+}^2}{m_{H^+}^2} \right)~.
\end{equation}
Demanding $m_{\nu_\tau} \leq 0.05$ eV, while also requiring $\mu_{\nu_\mu\nu_e} \sim 1.6 \times 10^{-11}\mu_B$ one obtains a limit $|h_3| < 10^{-2}$.  Consequently, the neutral $\phi_1^0$ exchange of Fig. \ref{feynman} to the muon anomalous magnetic moment is negligible.  Thus we arrive at the model prediction that $\Delta a_\mu$ receives significant contributions only from $\phi_2^0$ exchange diagram.

In Fig.~\ref{plota} we show the parameter space in the Yukawa coupling $|f'|$ versus the mass of the neutral scalar ($m_{\phi^0}$) plane consistent with {the observed muon anomalous magnetic moment measurement $\Delta a_\mu= (251 \pm 59)\times 10^{-11}$ at Fermilab \cite{Abi:2021gix} as indicated by green band.} Here we have fixed the charged scalar masses at $m_{h^+} = m_{\phi^0} + 50$ GeV and $m_{H^+} = m_{h^+} + 350$ GeV, a choice consistent with electroweak $T$ parameter constraint. A lower bound on the neutral scalar mass is obtained from the decay width measurements of the $Z$ boson: $m_{\phi^0}\gtrsim 45$ GeV \cite{LEP:2003aa,Tanabashi:2018oca}, which is indicated by the yellow shaded region. The Yukawa coupling $f'$ leads to additional contributions to the $e^+ e^- \to \tau^+ \tau^- $ process at the LEP experiment \cite{LEP:2003aa} via $t-$channel neutral scalar exchange. This leads to an upper limit on  $f'$ which is depicted by the pink shaded region. Moreover, our model leads to the  most promising signal  $p p \rightarrow \phi_i^0 \phi_i^{*0 } \to \mu^{-} \mu^{+} \tau^{-} \tau^{+}$ at the LHC  where different flavored leptons can be easily reconstructed as a resonance. The collider phenomenology of this model has been studied in detail in Ref. \cite{Babu:2020ivd}.  The blue dashed line represents the 5$\sigma$ sensitivity of the neutral scalar of mass 398 GeV \cite{Babu:2020ivd} at the HL-LHC with an integrated luminosity of $\mathcal{L} =1$ ab$^{-1}$. This can be a promising test of the model for neutrino magnetic moment and muon $g-2$. We also note that the model predicts a deviation in the anomalous magnetic moment of the electron, $a_e$, which is calculable owing to the $SU(2)_H$ symmetry. We find it to be $\Delta a_e = \Delta a_\mu (m_e/m_\mu)^2 \approx 6 \times 10^{-14}$. This predicted deviation is consistent with the current measurements \cite{Parker_2018,Morel:2020dww}, but may serve as a future test of the model.

\begin{figure*}[htb!]
\includegraphics[width=0.7\textwidth]{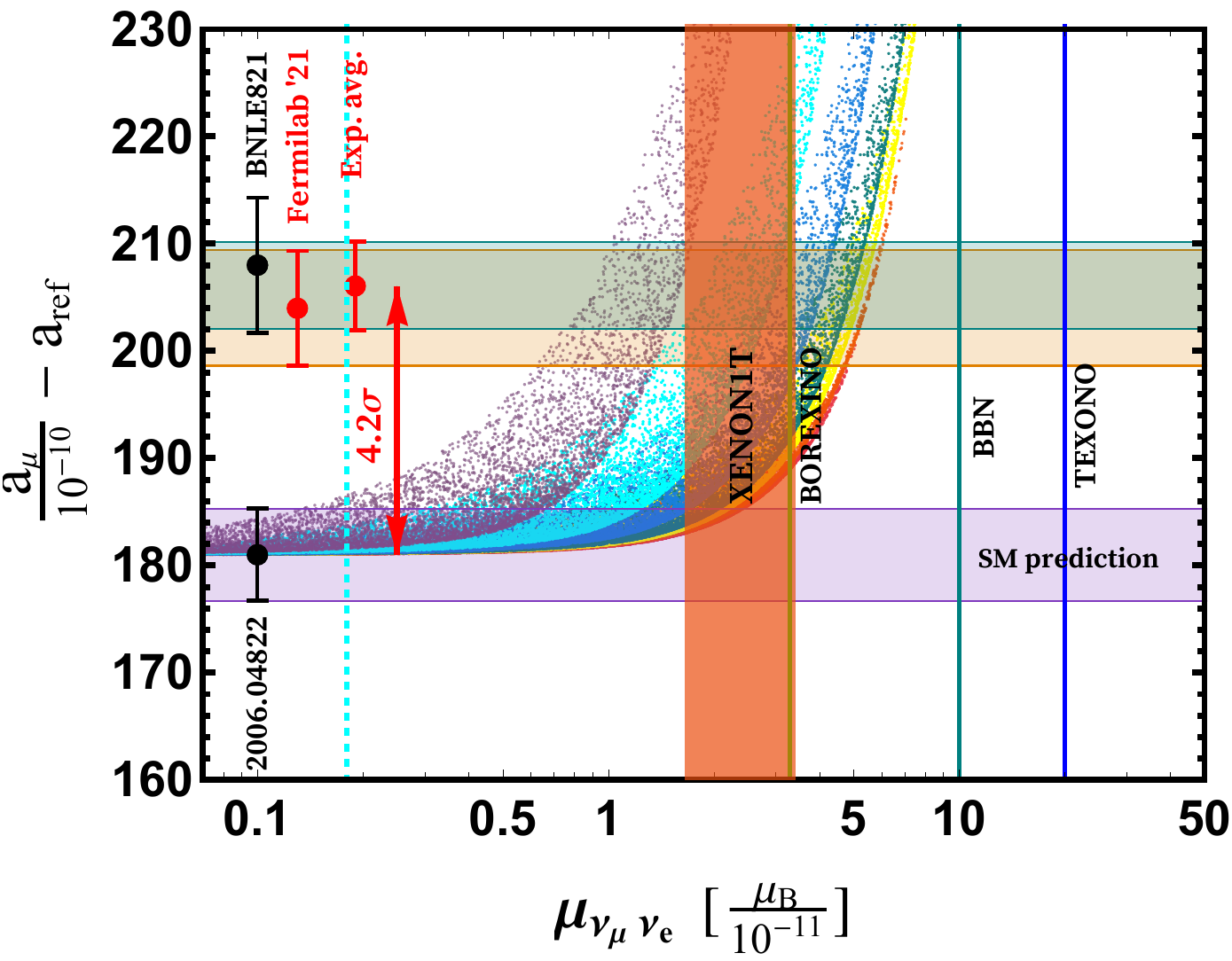}
\caption{Theoretical predictions  and experimental measurements of the muon anomalous magnetic moment and the neutrino transition magnetic moment. Purple, cyan, light blue, blue, green, yellow, orange and red  scattered points depict the correlated predictions for the muon magnetic moment ($a_{\mu}$) and neutrino magnetic moment ($\mu_{\nu_\mu \nu_e}$) in our framework for different ranges of mass splittings between the charged scalars: $m_{H^+}-m_{h^+}=20,50,100,200,300,500,800$ and 1000 GeV. Here we choose different ranges of parameters as $\lbrace f' \in (0.001-1), ~ m_{\phi^0} \in (45-500)$ GeV, and  $m_{H^+}-m_{h^+} \in (20$ GeV - 1 TeV)$\rbrace$ consistent with the theoretical and experimental limits discussed in text. The lower black dotted point with error bar indicates the updated SM prediction \cite{Aoyama:2020ynm} for  $a_{\mu}$ (see other previous analyses  (in chronological order) as: HMNT06 \cite{Hagiwara:2006jt}, DHMZ10 \cite{Davier:2010nc}, JS11 \cite{Jegerlehner:2011ti}, HLMNT11 \cite{Hagiwara:2011af}, DHMZ17 \cite{Davier:2017zfy}, KNT18 \cite{Keshavarzi:2018mgv}) , and  others \cite{Colangelo:2018mtw,Hoferichter:2019mqg,Davier:2019can,Keshavarzi:2019abf,Kurz:2014wya,Melnikov:2003xd,Masjuan:2017tvw,Colangelo:2017fiz,Hoferichter:2018kwz,Gerardin:2019vio,Bijnens:2019ghy,Colangelo:2019uex,Colangelo:2014qya,Blum:2019ugy,Aoyama:2012wk,Czarnecki:2002nt,Gnendiger:2013pva}).  The upper dotted data points with error bars represent different   experimental measurements: BNLE821 \cite{Bennett:2006fi}, {Fermilab \cite{Abi:2021gix}},  {and the the combined experimental average of the BNL \cite{Bennett:2006fi} and Fermilab \cite{Abi:2021gix} results}. These evaluations and measurements of $a_{\mu}$ are independent of  neutrino magnetic moment values.  {The current uncertainty on the measurement of $a_{\mu}$ at the Fermilab \cite{Abi:2021gix} is given by the light brown band at $1\sigma $ level}. The purple band represents the SM prediction from Ref.~\cite{Aoyama:2020ynm} with $1\sigma$ value and { the light green band indicates the combined experimental average of the BNL \cite{Bennett:2006fi} and Fermilab \cite{Abi:2021gix} results  at $1\sigma $ level}. Here we choose the reference value $a_{ref} = 11659000.$ The orange band labelled with XENON1T denotes the preferred range $\mu_{\nu_{e} \nu_{\mu}} \in(1.65-3.42) \times 10^{-11} \mu_{B}$ to explain the  electron recoil excess at $90\%$ C.L \cite{Aprile:2020tmw}. The vertical lines indicate the limits on neutrino magnetic moment from various measurements: blue solid line from TEXONO \cite{Deniz:2009mu}, dark green solid line from BBN \cite{Vassh:2015yza}, light green solid line from BOREXINO \cite{Borexino:2017fbd}, and cyan dashed line from globular clusters \cite{Viaux:2013lha}. The astrophysical limits on neutrino magnetic moment shown by the dashed cyan line  can be evaded by utilizing a neutrino trapping mechanism \cite{Babu:2020ivd} (see discussions in text).
} \label{main2}
\end{figure*}

In Fig.~\ref{main2} we have shown a direct correlation between the muon anomalous magnetic moment and neutrino magnetic moment within our framework.  As noted before, the dominant contribution to $\Delta a_\mu$ solely depends on the Yukawa coupling $f'$ and the neutral scalar mass $m_{\phi^0}$. The transition magnetic moment of the neutrino $\mu_{\nu_\mu\nu_e}$  depends on the same Yukawa coupling $f'$ as well as the charged scalar masses. The measurement of the  $e^+ e^- \to \tau^+ \tau^-$ process at the LEP experiment \cite{LEP:2003aa} imposes a strong limit on the Yukawa coupling $f'$  as a function of the neutral scalar mass $m_{\phi_0}$.  Therefore, in Fig.~\ref{main2}, we vary the Yukawa coupling $f'$  in such a way that the parameters are consistent with these constraints. An optimal scenario is realized when the neutral scalar is light, with its mass not lower than 45 GeV so as to be consistent with $Z$ decay constraint \cite{Tanabashi:2018oca}. The mass splitting between the lightest charged scalar and neutral scalar is set to be 50 GeV in order to satisfy the charged Higgs mass limit of $\sim$95 GeV (for detail see Ref. \cite{Babu:2019mfe}) from collider searches. As for the charged scalar masses which have a strong impact on the neutrino magnetic moment, but little effect on the muon $g-2$, we note that these masses cannot be arbitrary, as the mass splittings between the charged and neutral scalars are  tightly bounded  from the electroweak precision data \cite{Tanabashi:2018oca, Babu:2019mfe}.  Therefore, it follows that  there is no extra room to control the strength and the sign of the muon $g-2$ in our setup. The parameter space which predicts large neutrino magnetic moment leads to {\it a positive} contribution to muon $g-2$.  {Thus, the measurement of muon $g-2$ by the Fermilab experiment can be an indirect and novel test of the neutrino magnetic moment hypothesis, which can be as sensitive as other ongoing neutrino/dark matter experiments.}

There are strong astrophysical constraints on neutrino magnetic moments from red giants and horizontal branch stars, since  photons in the plasma of these environments can decay into neutrino pairs \cite{Bernstein:1963qh,Raffelt:1999tx}.  A limit $|\mu_\nu| \leq 1.5 \times 10^{-12} \mu_B$  (95\% CL) has been derived from the evolutionary studies of horizontal branch stars \cite{Viaux:2013lha, Viaux:2013hca,  Capozzi:2020cbu}. This would be in contradiction with the $\mu_{\nu_e\nu_\mu}$ needed to explain the XENON1T anomaly. However, this constraint can be evaded if $\nu_\mu$ has  a matter-dependent mass,  which can arise from its weak coupling to a new light scalar, as shown in Ref. \cite{Babu:2019iml} and adopted in Ref. \cite{Babu:2020ivd}.  Note that the transition magnetic moment $\mu_{\nu_e \nu_\mu}$ would lead to plasmon decay into $\nu_e + \nu_\mu$, which would be kinematically suppressed inside red giants and horizontal branch stars if the matter-dependent mass of $\nu_\mu$ is of order few keV. This would not suppress the standard $\nu_e$ emission, which could have modified the evolution of these stars significantly \cite{Ahlgren:2013wba,  Straniero:2020iyi}. The modifications to neutrino emissivity will only affect $\nu_\mu \overline{\nu}_\mu$  emissions mediated through neutral currents.  Following Ref. \cite{Braaten:1993jw}, we estimate that the neutrino emissivity changes by only $\sim0.3\%$ in our scenario compared to the standard case. This shift is small owing to the dominance of the vector-couplings of the electrons to the $Z$ boson, which is however suppressed by a factor $(1-4 \sin^2\theta_W)$. The contribution of the axial-vector current is always negligible for all conditions of astrophysical interest \cite{1986ApJ...310..815K,Raffelt:1996wa}.

{\textbf {\textit {Conclusions.--}}} 
In this paper we have analyzed new contributions to the muon anomalous magnetic moment in 
a class of models that generates naturally large transition magnetic moment for the neutrino needed to explain the XENON1T electron recoil excess.  These models are based on an approximate $SU(2)_H$ symmetry that suppresses the neutrino mass while allowing for a large neutrino transition magnetic moment. We have shown that the new scalars present in the theory with masses around 100 GeV can yield the right sign and magnitude for the muon $g-2$  which has been confirmed recently by the Fermilab collaboration. \color{black} Such a correlation between muon $g-2$ and the neutrino magnetic moment is generic in models employing leptonic family symmetry to explain a naturally  large $\mu_{\nu_\mu \nu_e}$. We have also outlined various other experimental tests of these models at colliders. The entire parameter space of the model can be explored at the HL-LHC through the pair production of neutral scalars and their subsequent decays into $e\tau$ and $\mu\tau$ final states.  
\vspace{0.1in}
\begin{acknowledgments}
{\textbf {\textit {Acknowledgments.--}}} We thank Evgeny Akhmedov for discussions. The work of KSB and VPK is in part supported by US Department of Energy Grant Number DE-SC 0016013 and by a Fermilab Neutrino Theory Network grant.
 \end{acknowledgments}

\vspace{-0.398in}
\bibliographystyle{utphys}
\bibliography{reference}

\end{document}